\documentclass[11pt,rmp,twocolumn,natbib,showkeys,amsmath] {revtex4}

\usepackage[dvips]{graphicx}
\begin{document}


\title{Volatility Effects on the Escape Time in Financial Market Models}

\author{Bernardo Spagnolo$^{\star}$ and Davide Valenti$^{\circ}$}%
 \affiliation{Dipartimento di Fisica e Tecnologie Relative,
  Group of Interdisciplinary
  Physics\footnote {Electronic address: http://gip.dft.unipa.it},
  Universit\`a di Palermo,
 \\
Viale delle Scienze pad.~18, I-90128 Palermo, Italy
\\$^{\star}$spagnolo@unipa.it,
$^{\circ}$valentid@gip.dft.unipa.it}
\date{\today}

\begin{abstract}
We shortly review the statistical properties of the escape times, or
hitting times, for stock price returns by using different models
which describe the stock market evolution. We compare the
probability function (PF) of these escape times with that obtained
from real market data. Afterwards we analyze in detail the effect
both of noise and different initial conditions on the escape time in
a market model with stochastic volatility and a cubic nonlinearity.
For this model we compare the PF of the stock price returns, the PF
of the volatility and the return correlation with the same
statistical characteristics obtained from real market data.
\end{abstract}

\pacs{89.65.Gh; 02.50.-r; 05.40.-a; 89.75.-k}
\keywords{Econophysics, Stock market model, Langevin-type equation,
Heston model, Complex Systems}
\maketitle

\section{\label{sec:intro}Introduction}
\vskip-0.4cm

Econophysics is a developing interdisciplinary research field in
recent years. It applies theories and methods originally developed
by physicists in statistical physics and complexity in order to
solve problems in economics such as those strictly related to the
analysis of financial market data [Anderson \emph{et al.}, 1988,
1997; Mantegna \& Stanley, 2000; Bouchaud \& Potters, 2004]. Most of
the work in econophysics has been focused on empirical studies of
different phenomena to discover some universal laws. Recently more
effort has been done to construct new models. In a real market the
stock option evolution is determined by many traders which interact
with each other and use different strategy to increase their own
profit. The market is then "pushed" by many different forces, which
often affect the system in such a way that every deterministic
forecast is impossible. In fact people act in the market so that any
forecast results to be unpredictable. The arbitrariness of each
choice, together with the non-linearity of the system, leads to
consider the stock option market as a complex system where the
randomness of the human behavior can be modelled by using stochastic processes.\\
\indent For decades the geometric Brownian motion, proposed by Black
and Scholes [Black \& Scholes, 1973] to address quantitatively the
problem of option prices, was widely accepted as one of the most
universal models for speculative markets. However, it is not
adequate to correctly describe financial market behavior [Mantegna
\& Stanley, 2000; Bouchaud \& Potters, 2004]. A correction to
Black-Scholes model has been proposed by introducing stochastic
volatility models. These models are used in the field of
quantitative finance to evaluate derivative securities, such as
options, and are based on a category of stochastic processes that
have stochastic second moments. In finance, two categories of
stochastic processes are widely used to model stochastic second
moments. One is represented by the stochastic volatility models
(SVMs), the other one by ARCH/GARCH models [Engle, 1982; Bollerslev,
1986], where the present volatility depends on the past values of
the square return (ARCH) and also on the past values of the
volatility (GARCH). Both ARCH/GARCH and stochastic volatility models
derive their randomness from white noise processes. The difference
is that an ARCH/GARCH process depends on just one white noise, while
SVMs generally depend on two white noises and they model the
tendency of volatility to revert to some long-run mean value.
Stochastic volatility models address many of the short-comings of
popular option pricing models such as the Black-Scholes model [Black
\& Scholes, 1973] and the Cox-Ingersoll-Ross (CIR) model [Cox
\emph{et al.}, 1985]. In particular, these models assume that the
underlier volatility is constant over the life of the derivative,
and unaffected by the changes in the price level of the underlier.
However, these models can not explain long-observed anomalies such
as the \emph{volatility smile} [Fouque \emph{et al.}, 2000] and some
stylized facts observed in financial time series such as long range
memory and clustering of the volatility, which indicate that
volatility does tend to vary over the time [Dacorogna \emph{et al.},
2001]. By assuming that the volatility of the underlying price is a
stochastic process rather than a constant parameter, it becomes
possible to more accurately model derivatives. In SVMs the
volatility is changing randomly according to some stochastic
differential equation or some discrete random processes. Recently,
models of financial markets reproducing the most prominent
statistical properties of stock market data, whose dynamics is
governed by non-linear stochastic differential equations, have been
proposed [Malcai \emph{et al.}, 2002; Borland, 2002; Borland, 2002b;
Hatchett \& K$\ddot{u}$hn, 2006; Bouchaud \& Cont, 1998; Bouchaud,
2001; Bouchaud, 2002; Sornette, 2003; Bonanno \emph{et al.}, 2006;
Bonanno \emph{et al.}, 2007].\\ \indent In particular some models
have been used where the market dynamics is governed, close to
crisis period, by a cubic potential with a metastable state
[Bouchaud \& Potters, 2004; Bouchaud \& Cont, 1998; Bouchaud, 2001;
Bouchaud, 2002; Bonanno \emph{et al.}, 2006; Bonanno \emph{et al.},
2007]. The metastable state is connected with the stability of
normal days, when the financial market shows a normal behavior.
Conversely the presence of a crisis is modelled as an escape event
from the metastable state and the subsequent trajectory. The
importance of metastable states in real systems, ranging from
biology, chemistry, ecology to population dynamics, social sciences,
economics, caused researchers to devote many efforts to investigate
the dynamics of metastable systems, finding that they can be
stabilized by the presence of suitable levels of noise intensity
[Mantegna \& Spagnolo, 1996; Mielke, 2000; Agudov \& Spagnolo, 2001;
Dubkov \emph{et al.}, 2004; Fiasconaro \emph{et al.}, 2005;
Fiasconaro \emph{et al.}, 2006; Bonanno \emph{et al.}, 2007].\\
\indent Our focus in this paper is to analyze the statistical
properties of the escape times in different market models, by
comparing the probability function (PF) with that observed in real
market data. Recent work has been done on the mean exit time
[Bonanno \& Spagnolo, 2005; Montero \emph{et al.}, 2005] and the
waiting time distribution in financial time series [Raberto \emph{et
al.}, 2002]. Here, starting from the geometric random walk model we
shortly review the statistical properties of the escape times for
stock price returns in some stochastic volatility models as GARCH,
Heston and nonlinear Heston models. In the last model, recently
proposed by the authors [Bonanno \emph{et al.}, 2006; Bonanno
\emph{et al.}, 2007] and characterized by a cubic nonlinearity, we
compare some of the main statistical characteristics, that is the PF
of the stock price returns, the PF of the volatility and the return
correlation, with the same quantities obtained from real market
data. We also analyze in detail the effect of the noise and
different initial conditions on the escape times in this nonlinear
Heston model (NLH).

\section{Escape times in stock market models}
\vskip-0.4cm

\begin{figure}[htbp]
\vspace{5mm}
\centering{\resizebox{8cm}{!}{\includegraphics{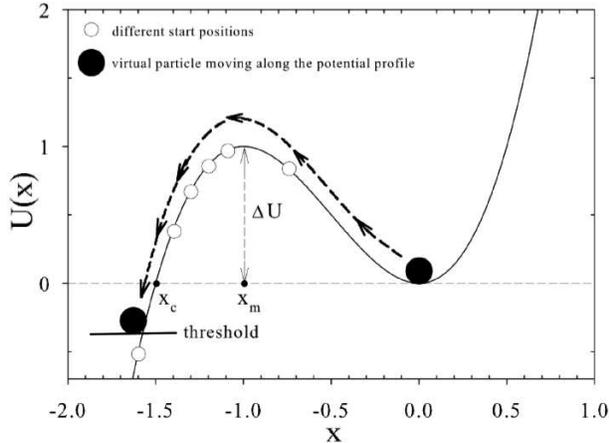}}}
\caption{\label{Fig:Cubico} Cubic potential used in the dynamical
equation for the process x(t) of Eq.~(7). The white circles in the
figure indicate the starting positions used in our simulations.}
\end{figure}

The average escape time of a Brownian particle, moving in a
potential profile, is a well-known problem in Physics [Gardiner,
2004]. This quantity is defined as $<T>\thinspace=\thinspace
\int_0^\infty f(t) dt$, where $f(t)$ is the probability function of
the escape events from a certain region of the potential (see
Fig.~\ref{Fig:Cubico}).
In other words, the average escape time is the mean time that a
particle, starting from a certain initial point, takes to pass
through a given threshold. This average time, obtained by using the
backward Fokker-Planck equation, corresponds to the first passage
time used in statistical physics [Redner, 2001; Inoue \& Sazuka,
2007] and to the first hitting time defined in econophysics [Bonanno
\& Spagnolo, 2005; Montero \emph{et al.}, 2005].

\subsection{The geometric Brownian motion model}
\vskip-0.4cm

A common starting point for many theories in economics and finance
is that the stock price, in the continuous limit, is a stochastic
multiplicative process defined, in the Ito sense, as
\vskip-0.7cm
\begin{equation}
  d\,p(t) = \gamma \cdot p(t) \cdot dt + \sigma \cdot p(t) \cdot dW(t)
\label{Multiplicative}
\end{equation}
where $\gamma$ and $\sigma$ are, in the market dynamics, the
expected average growth for the price and the expected noise
intensity ({\it volatility}) respectively. The price return $dp/p =
d\,lnp(t)$ obeys the following additive stochastic differential
equation
\begin{equation}
  d\,lnp(t) = (\mu - \frac{\sigma^2}{2}) \cdot dt + \sigma \cdot
  dW(t)\;.
\label{Additive}
\end{equation}
This simple market model, proposed by Black and Scholes, catches one
of the most important stylized facts of the financial markets, that
is the short range correlation for price returns. This
characteristic is necessary in order to warrant market efficiency.

In the geometric Brownian motion the returns are independent on each
other, so the probability to observe a value after a certain barrier
is given by the probability that the "particle" doesn't escape after
$n-1$ time steps, multiplied by the escape probability at the
$n^{th}$ step
\begin{eqnarray}
 F(\tau) &=& (1-p) \cdot p^{n-1}\\
 &=& (1-p) \cdot exp\left[(n-1)\ln p
 \right], \;\;\; n = \frac{\tau}{\Delta t}\nonumber
 \end{eqnarray}
where $p$ is the probability to observe a return inside the region
limited by the barrier, $\Delta t$ is the observation time step and
$\tau$ is the escape time. So the behavior of the PF of hitting
times is exponential. The geometric Brownian motion however is not
adequate to describe financial market behavior. In fact the
volatility is considered as a constant parameter and the PF of the
price is a log-normal distribution. As a consequence many
observations of real data are in clear disagreement with this model
[Mantegna \& Stanley, 2000; Bouchaud \& Potters, 2004].

\subsection{Stochastic volatility models}
\subsubsection{The GARCH model}

Data on financial return volatility are influenced by time dependent
information flow which results in pronounced temporal volatility
clustering. These time series can be parameterized using Generalized
Autoregressive Conditional Heteroskedastic (GARCH) models. It has
been found that GARCH models can provide good in-sample parameter
estimates and, when the appropriate volatility measure is used,
reliable out-of-sample volatility forecasts [Anderson \& Bollerslev,
1998].

The GARCH(p,q) process, which is essentially a random multiplicative
process, is the generalization of the ARCH  process and combines
linearly the present return with the $p$ previous values of the
variance and the $q$ previous values of the square return
[Bollerslev, 1986]. The process is described by the equations
\vskip-0.4cm
\begin{eqnarray}
  \sigma ^2 _t &=& \alpha_0 + \alpha_1x ^2 _{t-1} + \dots +
  \alpha ^2 _{q} x ^2 _{t-q} +
  \beta_1 \sigma ^2 _{t-1} \nonumber \\
  &+& \dots + \beta_p \sigma ^2 _{t-p},
  \;\;\;\;\;\;\;
   x_t = \eta _t \cdot \sigma _t,
\label{garch}
\end{eqnarray}
where $\alpha_i$ and $\beta_i$ are parameters that can be estimated
by means of a best fit of real market data, $\eta_t$ is an
independent identically distributed random process with zero mean
and unit variance. Using the assumption of Guassian conditional PF,
$\eta_t$ is Gaussian. In Eq.(\ref{garch}) $x_t$ is a stochastic
process representing price returns and is characterized by a
standard deviation $\sigma_t$. The GARCH process has a non-constant
conditional variance but the variance observed in long time period,
called unconditional variance, is instead constant and can be
calculated as a function of the model parameters. We shall consider
the simpler GARCH(1,1) model
\begin{equation}
  \sigma ^2 _t = \alpha_0 + (\alpha_1\eta_{t-1}^2 + \beta_1)\sigma ^2 _{t-1},
  \;\;
  x_t = \eta _t \cdot \sigma _t \;.
   \label{garch11}
\end{equation}
The autocorrelation function of the process $x_t$ is proportional
to a delta function, while the process $x_t^2$ has a correlation
characteristic time equal to $\tau = \mid ln(\alpha_1+\beta_1)
\mid^{-1}$ and the unconditional variance equal to $\sigma ^2 =
\alpha_0 /(1- \alpha_1 - \beta_1)$. By a fitting procedure between
the previous expressions of $\tau$ and $\sigma ^2$ and the
empirical values of the same quantities, we can easily estimate
the three parameters $\alpha_0$, $\alpha_1$ and $\beta_1$ which
characterize the model.

\subsubsection{The Heston model}

The Heston model introduced by Heston [Heston, 1993] is a commonly
used stochastic volatility model. It received a great attention in
the financial literature specially in connection with option pricing
[Fouque \emph{et al.}, 2000]. The Heston model was verified
empirically with both stocks [Silva \& Yakovenko, 2003,
Dr$\check{a}$gulescu \& Yakovenko, 2002] and options [Hull \& White,
1987; Hull, 2004], and good agreement with the data has been found.
It was also recently investigated by econophysicists [Miccich\`e
\emph{et al.}, 2002; Dr$\check{a}$gulescu \& Yakovenko, 2002; Silva
\emph{et al.}, 2004; Bonanno \& Spagnolo, 2005]. The model is
defined by two coupled stochastic differential equations which
represent the stock dynamics by the log-normal geometric Brownian
motion stock process and the Cox-Ingersoll-Ross (CIR) mean-reverting
process (SDE), first introduced to model the short term interest
rate [Cox \emph{et al.}, 1985]. By considering the \emph{log} of the
price $x(t) = ln~p(t)$ the SDEs of the model are
\begin{eqnarray}
  dx(t) & = & \left(\mu - \frac{v(t)}{2}\right) \cdot dt + \sqrt{v(t)} \cdot dW_1(t) \nonumber \\
  dv(t)   & = & a(b-v(t)) \cdot dt + c \sqrt{v(t)} \cdot dW_c(t) \nonumber \\
  dW_c(t) & = & \rho \cdot dW_1(t) + \sqrt{1-\rho^2} \cdot dW_2(t),\;\;
\label{heston}
\end{eqnarray}
where $\mu$ is the trend of the market, $W_1(t)$ and $W_2(t)$ are
uncorrelated Wiener processes with the usual statistical properties
$<dW_i>\thinspace = \thinspace 0, \;$ $<dW_i(t)dW_j(t')>\thinspace =
\thinspace dt~\delta(t-t')~\delta_{i,j}$ ($i,j=1, 2$), and $\rho$ is
the cross correlation coefficient between the noise sources. Here
$v$ is the CIR process, which is defined by three parameters: $b$,
$a$ and $c$. They represent respectively the long term variance, the
rate of mean reversion to the long term variance, and the volatility
of variance, often called the \emph{volatility of volatility}. The
stochastic volatility $v(t)$ is characterized by exponential
autocorrelation and \emph{volatility clustering} [Cont, 2001;
Bouchaud \& Potters,2004; Bonanno \emph{et al.}, 2006], that is
alternating calm with burst periods of volatility.\\
\indent We end this paragraph comparing the PF of the escape times
$\tau$ of the price returns obtained from real market data with the
PFs obtained by using the three previous models. We use a set of
returns obtained from the daily closure prices for $1071$ stocks
traded at the NYSE and continuously present in the $12-$year period
$1987-1998$ (3030 trading days). From this data set we obtain the
time series of the returns and we calculate the time to hit a fixed
threshold starting from a fixed initial position. The parameters in
the models were chosen by means of a best fit, in order to reproduce
the correlation properties and the variance appropriate for real
market [Bonanno \emph{et al.}, 2005]. We choose two thresholds to
define the start and the end for the random walk. Specifically we
calculate the standard deviation $\sigma_i$, with $i=1,\dots,1071$
for each stock over the whole 12-year period. Then we set the
initial threshold at the value $-0.1 \cdot \sigma_i$ and as final
threshold the value $-2 \cdot \sigma_i$. The thresholds are
different for each stock, the final threshold is considered as an
absorbing barrier. The results of our simulations together with the
real market data are shown in the following Fig.~\ref{taudistro}. In
this figure the exponential behavior represents the PF of the escape
times for the geometric Brownian motion model, which is not adequate
do describe correctly the PF of $\tau$ over the entire time axis.
The GARCH model provides a better qualitative agreement with real
data for lower escape times and gives the exponential behaviour in
the region of large escape times. Here the geometric Brownian model
reproduces well the real data\footnote{By changing the fit
parameters $\alpha_1$ and $\beta_1$ for the GARCH model it is
possible to obtain a better good agreement with the real data.},
whereas the Heston model is able to reproduce almost entirely the
empirical PF.
\begin{figure}[htbp]
\vspace{5mm}
\centering{\resizebox{7.5cm}{!}{\includegraphics{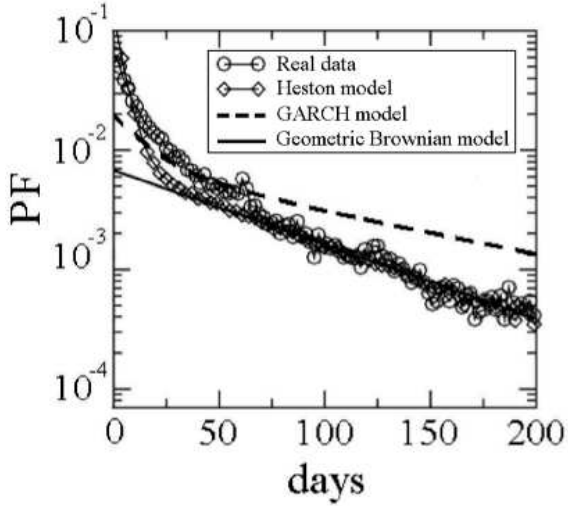}}}
\caption{Probability function (PF) of escape times of stock price
returns for: real market data (circle), geometric Brownian motion
model (black solid line), GARCH model (black broken line) and Heston
model (diamond).} \label{taudistro}
\end{figure}

\subsection{The nonlinear Heston model (NLH)}

\begin{figure}[htbp]
\vspace{5mm}
\centering{\resizebox{7.5cm}{!}{\includegraphics{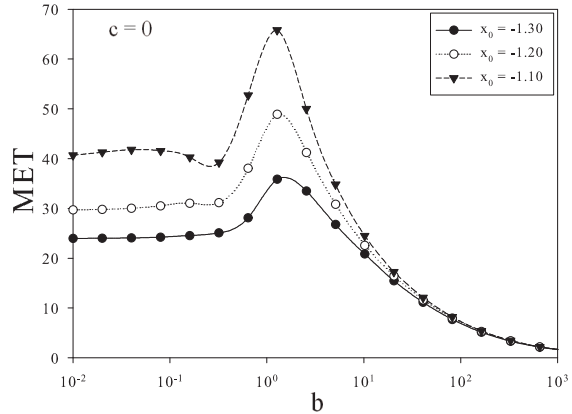}}}
\vskip-0.2cm \caption{\label{revertlimit} Mean escape time (MET) for
$3$ different unstable starting positions, when only the reverting
term is present: $a = 10^{-2}$, $c = 0$. The curves are averaged
over $10^5$ escape events.}
\end{figure}

To consider feedback effects on the price fluctuations and different
dynamical regimes, similarly to what happens in financial markets
during normal activity and in special days with relatively strong
variations of the price [Bouchaud \& Potters, 2004; Bouchaud \&
Cont, 1998; Bouchaud, 2001; Bouchaud, 2002; Bonanno \emph{et al.},
2006; Bonanno \emph{et al.}, 2007], we proposed a generalization of
the Heston model, by considering a cubic nonlinearity in the SDE of
the $log$ of the price $x(t) = ln~p(t)$ (first equation in (6))
[Bonanno \emph{et al.}, 2006; Bonanno \emph{et al.}, 2007]. This
nonlinearity allows us to describe these different dynamical regimes
by the motion of a fictitious "\emph{Brownian particle}" moving in
an \emph{effective} potential with a metastable state. The equations
of the new model are obtained by replacing in Eqs.~(\ref{heston})
the parameter $\mu$ with the negative derivative of the nonlinear
cubic potential
\begin{eqnarray}
  dx(t)   & = & - \left(\frac{\partial U}{\partial x} +
  \frac{v(t)}{2}\right) dt + \sqrt{v(t)} dW_1(t)\label{Eqn:NLH_1}\;\;\;\;\;  \\
  dv(t)   & = & a(b-v(t)) dt + c \sqrt{v(t)} dW_c(t)\label{Eqn:NLH_2}  \\
  dW_c(t) & = & \rho dW_1(t) + \sqrt{1-\rho^2}
  dW_2(t)\label{Eqn:NLH_3},
\end{eqnarray}
where $U(x)= 2x^3+3x^2$ is the \emph{effective} cubic potential with
a metastable state at $x_{me} = 0$, a maximum at $x_m = -1$, and a
cross point between the potential and the $x$ axes at $x_c = -1.5$
(see Fig.~\ref{Fig:Cubico}). The average exit time of the system
from the stable to the unstable domain of the potential shown in
Fig.~\ref{Fig:Cubico} may be prolonged by imposing external noise:
this phenomenon is named noise enhanced stability (NES). The
stability of systems with a metastable state can be increased by
enhancing the lifetime of the metastable state or the average exit
time of the system from the well. The NES effect was experimentally
observed in a tunnel diode [Mantegna \& Spagnolo, 1996] and in an
underdamped Josephson junction [Sun \emph{et el.}, 2007] and
theoretically predicted in a wide variety of systems such as chaotic
map, Josephson junctions, neuronal dynamics models and tumor-immune
system models [Mielke, 2000; Agudov \& Spagnolo, 2001; Dubkov
\emph{et al.}, 2004; Pankratova \emph{et al.}, 2004; Pankratov \&
Spagnolo, 2004; Fiasconaro \emph{et al.}, 2005; Fiasconaro \emph{et
al.}, 2006]. Two different dynamical regimes are observed depending
on the initial position of the Brownian particle along the potential
profile. One is characterized by a nonmonotonic behavior of the
lifetime, as a function of the noise intensity (here the volatility
$v(t)$), for initial positions $x_o < x_c$. The other one features a
divergence of the lifetime when the noise tends to zero for initial
positions $x_c < x_o < x_m$, implying that the Brownian particle
remains trapped inside the metastable state in the limit of small
noise intensities. In this dynamical regime a nonmonotonic behavior
of the lifetime with a minimum and a maximum as a function of the
noise intensity is also observed. This trapping phenomenon is always
observable when initial unstable positions of the Brownian particle
are near a metatastable state of the system investigated. The NES
effect and its different dynamical regimes can be explained
considering the barrier "\emph{seen}" by the Brownian particle
starting at the initial position $x_o $, that is $\Delta U_{in} =
U(x_{max})-U(x_o)$, and by comparing it with the height of the
barrier $\Delta U$ characterizing the metastable state (see
Fig.~\ref{Fig:Cubico}) [Agudov \& Spagnolo, 2001; Fiasconaro
\emph{et al.}, 2005]. For example for unstable initial positions
such as $x_c < x_o < x_m$ we have $\Delta U_{in} < \Delta U$ and
from a probabilistic point of view, it is easier to enter into the
well than to escape from, once the particle is entered. So a small
amount of noise can increase the lifetime of the metastable state.
When the noise intensity $v$ is much greater than $\Delta U$, the
typical exponential behavior is recovered. \\ \indent By
investigating the mean escape time (MET), as a function of the model
parameters $a$, $b$ and $c$, we found the parameter region where a
nonmonotonic behavior of MET is observable in our NLH model with
stochastic volatility $v(t)$ [Bonanno \emph{et al.}, 2006; Bonanno
\emph{et al.}, 2007]. This behaviour is similar  to that observed
for MET~versus~$v$ in the NES effect with constant volatility $v$.
We call the enhancement of the mean escape time (MET), with a
nonmonotonic behavior as a function of the model parameters, NES
effect in a broad sense. Two limit regimes characterize our NLH
model, one corresponding to the case $a=0$, with only the noise term
in the equation for the volatility $v(t)$, and the other one
corresponding to the case $c=0$ with only the reverting term in the
same Eq.~(8). In the first case ($a=0$), the system becomes too
noisy and the NES effect is not observable in the behavior of MET as
a function of the parameter $c$. In the second case ($c=0$), after
an exponential transient, the volatility reaches the asymptotic
value $b$, and the NES effect is observable as a function of $b$.
This case corresponds to the usual constant volatility regime.\\
\indent By considering the two noise sources $W_1(t)$ and $W_c(t)$
of Eqs.~(7) and (8) completely uncorrelated ($\rho = 0$), the
results of simulations of the NLH model (Eqs.~(7)-(9)), in the
second case ($c =0$), are reported in Fig.~\ref{revertlimit}, where
MET versus $b$ is plotted for three different starting unstable
initial positions and for $c =0$.
The simulations were performed considering the initial positions of
the process $x(t)$ in the unstable region $[x_c,x_m]$ and using an
absorbing barrier at $x=-6.0$. When the process $x(t)$ hits the
barrier, the escape time is registered and another simulation is
started, placing the process at the same starting position $x_o$,
but using the volatility value of the barrier hitting time. The
nonmonotonic behavior, which is more evident for starting positions
near the maximum of the potential, is always present. After the
maximum, when the values of $b$ are much greater than the potential
barrier height, the exponential behavior is recovered. The results
of our simulations show that the NES effect can be observed as a
function of the volatility reverting level $b$, the effect being
modulated by the parameter $(ab)/c$. The phenomenon disappears if
the noise term is predominant in comparison with the reverting term.
Moreover the effect is no more observable if the parameter $c$
pushes the system towards a too noisy region.  When the noise term
is coupled to the reverting term, we observe the NES effect as a
function of the parameter $c$. The effect disappears if $b$ is so
high as to saturate the system.\\ \indent In financial markets the
\emph{log} of the price $x(t) = ln~p(t)$ and the volatility $v(t)$
can be correlated ($\rho \neq 0$), and a negative correlation
between the processes is known as \emph{leverage effect} [Fouque
\emph{et al.}, 2000]. A negative correlation between the logarithm
of the price and the volatility means that a decrease in $x(t)$
induces an increase in the volatility $v(t)$, and this causes the
Brownian particle to escape easily from the well. As a consequence
the mean lifetime of the metastable state decreases, even if the
nonmonotonic behavior is still observable. On the contrary, when the
correlation $\rho$ is positive, decrease in $x(t)$ indeed is
associated with decrease in the volatility, the Brownian particle
therefore stays more inside the well. The escape process becomes
slow and this increases further the lifetime of the metastable
state. The presence of correlation between the stochastic volatility
and the noise source which affects directly the dynamics of the
quantity $x(t) = \ln p(t)$ (as in usual market models) can influence
therefore the stability of the market. Specifically a positive
correlation between $x(t)$ and volatility $v(t)$ slows down the
walker escape process, that is it delays the crash phenomenon
increasing the stability of the market. Conversely a negative
correlation accelerates the escape process, lowering the stability
of the system [Bonanno \emph{et al.}, 2006].

\section{Role of the initial conditions and statistical features}

\begin{figure}[htbp]
\vspace{1mm}
\includegraphics*[height=12cm,width=8cm]{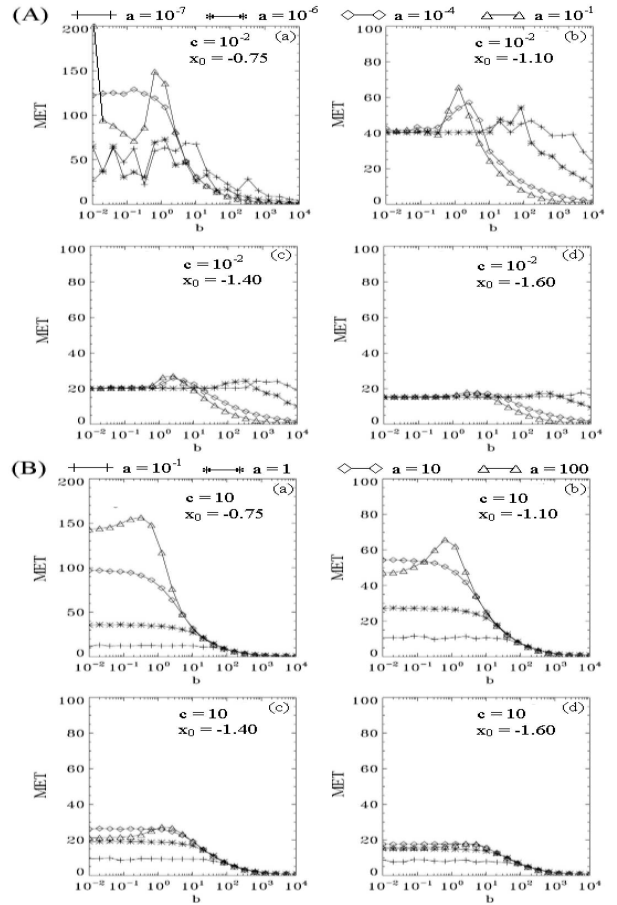}
 \vskip 0.0 cm \caption{\label{versusb} Mean escape time (MET) as a
function of reverting level $b$. (A) $c=10^{-2}$, (B) $c = 10$.
Each panel corresponds to different values of initial position.
Inside each panel different curves correspond to different values
of $a$.}
\end{figure}

\begin{figure}[htbp]
\vspace{1mm}
\centering{\resizebox{7.5cm}{!}{\includegraphics{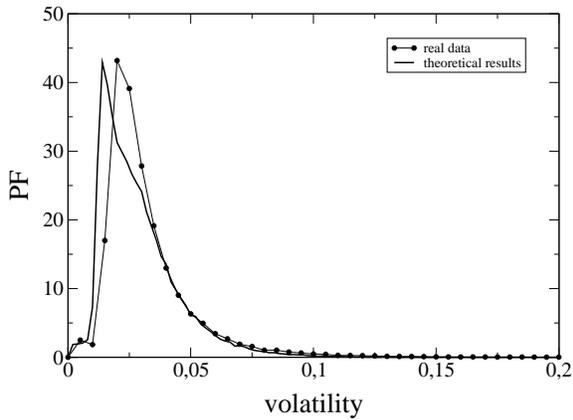}}}
\caption{Probability function of the volatility: (a) real data
(black circle), (b) NLH model (Eqs.~(7)-(9)) (solid line). The
values of the parameters are the same as in Fig.~\ref{pf return}.}
\label{pf volatility}
\end{figure}

\begin{figure}[htbp]
\vspace{1mm}
\includegraphics*[height=12cm,width=8cm]{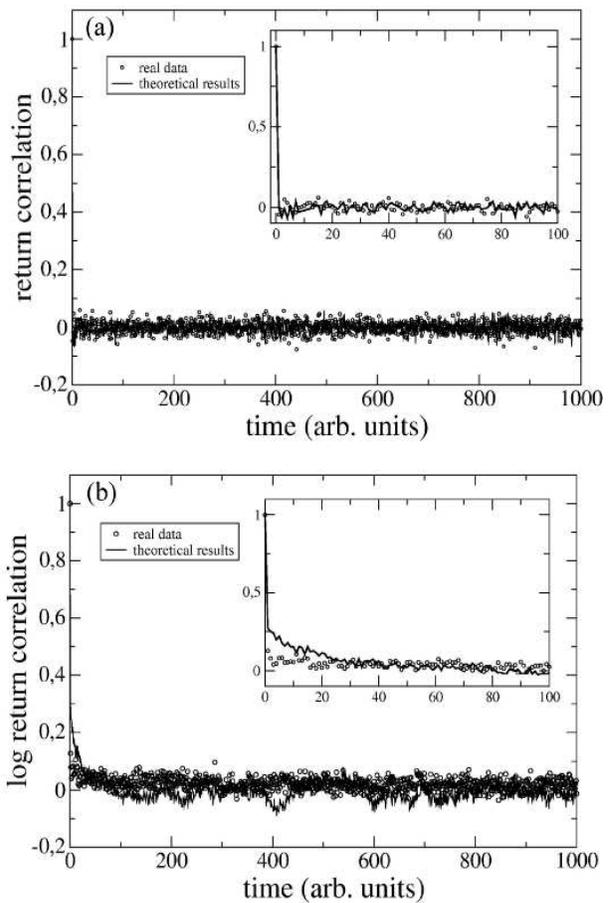}
\caption{(a) Correlation function of returns (panel a) and log
absolute returns (panel b): real data (circle), NLH model
(Eqs.~(\ref{Eqn:NLH_1})~-~(\ref{Eqn:NLH_3})) (solid line). Inset:
detail of the behaviour at short times. The values of the parameters
are the same of Fig.~\ref{pf return}. Inset: detail of the behaviour
at short times.} \label{return correlation}
\end{figure}

In this last section we study, for the uncorrelated ($\rho=0$) NHL
model (Eqs.~(7)-(9)), the role of the initial position of the
fictitious Brownian particle on the mean escape time (MET). In
particular we fix an escape barrier (threshold) and we analyze the
behaviour of MET as a function of $a$, $b$ and $c$ (CIR process
parameters) for different start positions (see
Fig~\ref{Fig:Cubico}). First we consider the behavior of MET as a
function of the reverting term $b$. In Fig.~\ref{versusb} (panel (a)
of part (A)) we show the curves averaged on $10^7$ escape events.
The curves inside all the other panels have been obtained averaging
on $10^5$ realizations. In our simulations we consider two different
values of the parameter $c$, namely $c = 10^{-2}, 10$, and eight
values of the parameter $a$, that is: (A) $a = 10^{-7}, 10^{-6},
10^{-4}, 10^{-1}$; and (B) $a = 10^{-1}, 1, 10, 10^{2}$. Each panel
corresponds to a different value of initial position, namely: (a)
$x_o = -0.75$, (b) $x_o = -1.10$, (c) $x_o = -1.40$, (d) $x_o =
-1.60$. Inside each panel different curves correspond to different
values of $a$. First of all we comment the panels (b), (c) and (d)
of part (A), related to unstable initial positions outside the
potential well. The nonmonotonic shape, characteristic of the NES
effect, is clearly shown in these three panels. This behavior is
shifted towards higher values of $b$ as the parameter $a$ decreases,
and it is always present. The NES effect is more pronounced for
initial positions near the top of the potential barrier.
For initial positions far from the maximum of the potential, the
trapping event becomes less probable. To obtain a more pronounced
NES effect we should consider very low values of the absorbing
barrier, that is $x \ll -6$, which are meaningless from financial
market point of view. For the higher value $c = 10$, we observe the
nonmonotonic behavior only for a very great value of the parameter
$a$, that is for $a = 100 \gg c$ (see panels (b), (c), (d) of
Fig.~\ref{versusb}B). For further increase of the parameter $c$ the
noise experienced by the system is much greater than the effective
potential barrier "\emph{seen}" by the fictitious Brownian particle
and the NES effect is never observable. We note that the parameters
$a$ and $c$ play a regulatory role in Eq.~(8). In fact for $a \gg c$
the drift term is predominant while for $a \ll c$ the dynamics is
driven by the noise term, unless the parameter $b$ takes great
values. The nonmonotonic behavior is observed for $a \ll c$,
provided that $b \gg c$. For increasing values of $a$ the system
approaches the revert-only regime and we recover the behavior shown
in Fig.~\ref{revertlimit}.\\ \indent Now we consider the panel of
Fig~\ref{versusb}A related to the unstable initial position outside
the potential well. For very low values of the parameter $a$ the
nonmonotonic behavior is absent and the mean escape time (MET)
decreases monotonically with strong fluctuations. We recover the
similar behavior obtained in the limit case of $a = 0$ and discussed
in [Bonanno \emph{et al.}, 2007]. For $a \ll c$, in fact, we can
neglect the reverting term in Eq.~(8) and the volatility is
proportional to the square of the Wiener process. The dynamics is
dominated by the noise term with large fluctuations for the MET.
This behavior is mainly due to the presence of the Ito term in
Eq.~(7) for log returns $x(t)$. The Ito term modifies randomly the
potential shape of Fig.~\ref{Fig:Cubico} in such a way that the
potential barrier disappears for greater values of the volatility
$v(t)$, producing a random enhancement of the escape process.
Increasing the value of $a$ these fluctuations disappear because the
reverting term becomes more important. This is the behavior shown
for $a = 10^{-4}$. A further increase of $a$ causes the revert term
to dominate the dynamics with respect to the noise term. Moreover,
because the initial unstable position $x_o = -0.75$ is near the
maximum of the potential well, we recover the divergent dynamical
regime characterized by a nonmonotonic behavior with a minimum and a
maximum of MET as a function of the noise intensity, here
represented by the parameter b [Fiasconaro \emph{et al.}, 2005]. For
very low values of $b$, the fictitious Brownian particle is trapped
inside the potential well with a divergence of MET in the limit $b
\rightarrow 0$. For increasing $b$ the particle can escape more
easily, and the MET decreases, as long as the noise intensity,
represented by the parameter $b$, reaches the value $0.15$
corresponding to the barrier height $\Delta U_{in}$ "\emph{seen}" by
the particle. Close to this value of $b$, the escape process is
slowed down, because the probability of reentering the well is equal
to that of escaping from. This behavior is represented by the
minimum of MET at $b \simeq 0.15$ in the panel (a) of
Fig.~\ref{versusb}A. By increasing $b$, the particle escaped from
the well can reenter as long as the noise intensity is comparable
with the height of the potential barrier. The MET therefore
increases until reach a maximum at $b \simeq \Delta U = 1$. At
higher values of $b$, one recovers a monotonic decreasing behavior
of the MET. The same nonmonotonic behavior with a minimum and a
maximum is visible in Fig.~\ref{revertlimit} for $x_o = -1.10$.
\begin{figure}[htbp]
\vspace{1mm}
\includegraphics*[height=6cm,width=8cm]{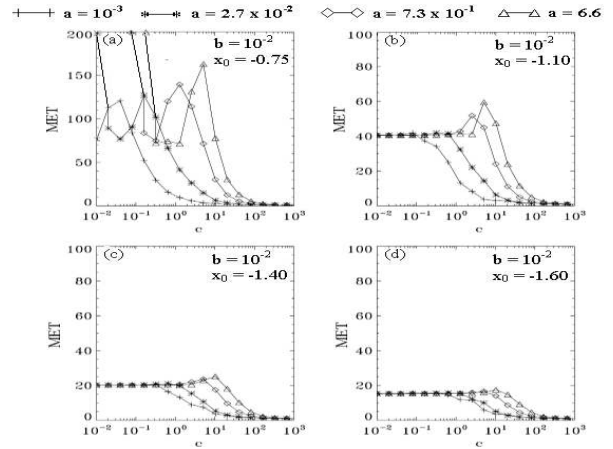}
\caption{\label{versusc} Mean escape time (MET) as a function of the
noise intensity $c$ for a fixed value of $b$ ($b = 10^{-2}$). Each
panel corresponds to different values of initial condition as in
Fig~\ref{versusb}. Inside each panel different curves correspond to
the following values of $a$: cross $10^{-3}$, star $2.7 \times
10^{-2}$, diamond $7.3 \times 10^{-1}$, triangle $6.6$. }
\end{figure}
Now we consider the dependence of MET on the noise intensity $c$.
Fig.~\ref{versusc} shows the curves of MET~versus~$c$, averaged over
$10^5$ escape events in panels (b), (c), (d) and $10^6$ escape
events in panel (a). First of all we consider the unstable initial
positions outside the well, that is panels (b), (c) and (d). Each
panel corresponds to a different value of initial position as in
Fig~\ref{versusb}. Inside each panel different curves correspond to
different values of $a$. The shape of the curves is similar to that
observed in Fig.~\ref{versusb}. Specifically for small values of
$a$, when the reverting term is negligible, the absence of the
nonmonotonic behavior is expected. By increasing $a$ the
nonmonotonic behavior is recovered. Again the NES effect is more
pronounced for initial positions near the maximum of the potential.
For initial position inside the potential well as in panel (a), we
observe a divergent behavior of MET for three values of the
parameter $a$ (namely $a = 2.7 \cdot 10^{-2}, 7.3 \cdot 10^{-1},
6.6$), because of the small value of $b = 10^{-2}$. Recall that the
volatility $v(t)$ reverts towards a long term mean squared
volatility $b$ with relaxation time given by $a^{-1}$. So, for
increasing values of $a$ the Brownian fictitious particle
experiences the low value of the noise intensity $b = 10^{-2}$ in a
shorter time and therefore the particle is trapped for relatively
small values of $c$ (see the curves for $a = 7.3 \cdot 10^{-1}$ and
$a = 6.6$). By decreasing the value of $a$, the relaxation time
increases considerably and the trapping of the particle occurs for
lower values of $c$ (see the curve for $a = 2.7 \cdot 10^{-2}$). For
the lowest value, $a = 10^{-3}$, we recover the regime of strong
fluctuations due to the predominance of the noise term with respect
to the revert term in Eq.~(8). The fluctuating behavior of all the
curves before the divergence in Fig.~\ref{versusc} is also due to
this effect.\\ \indent It is interesting to show, for our NLH model
(Eqs.~(7)-(9)), some of the well-established statistical features of
the financial time series, such as the probability function (PF) of
the stock price returns, the PF of the volatility and the return
correlation, and to compare them with the same characteristics
obtained from real market data. As initial position we choose
$x_o=-0.75$. For this start point, located inside the well, we have
very interesting behavior of the MET as a function of the noise
intensity $b$ (see panel (a) of Fig~\ref{versusb}A). In Fig.~\ref{pf
return} we show the PF of the returns. To characterize
quantitatively this PF with regard to the width, the asymmetry and
the fatness of the distribution, we calculate the mean value
$<\Delta x>$, the variance $\sigma_{\Delta x}$, the skewness
$\kappa_3$, and the kurtosis $\kappa_4$ for the NLH model and for
real market data. We obtain for \emph{real data}: $<\Delta x> =
0.221\cdot10^{-3}$, $\sigma_{\Delta x} = 0.0221$, $\kappa_3 = -
1.353$, $\kappa_4 = 79.334$; for the \emph{NLH model}: $<\Delta x> =
-1.909\cdot10^{-5}$, $\sigma_{\Delta x} = 0.0246$, $\kappa_3 = -
4.3501$, $\kappa_4 = 441.65$. The agreement between theoretical
results and real data is quite good except at high values of the
returns. These statistical quantities clearly show the asymmetry of
the distribution and its leptokurtic nature observed in real market
data. In fact, the empirical PF is characterized by a narrow and
large maximum, and fat tails in comparison with the Gaussian
distribution [Mantegna \& Stanley, 2000; Bouchaud \& Potters, 2004].
\begin{figure}[htbp]
\centering{\resizebox{7.5cm}{!}{\includegraphics{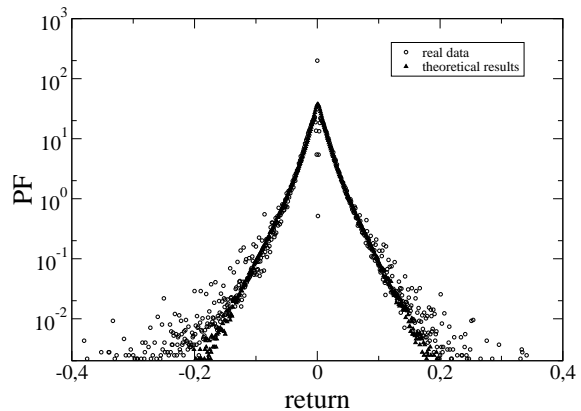}}}
\caption{Probability function of the stock price returns: (a) real
data (circle), (b) NLH model (Eqs.~(7)-(9)) (triangle). The values
of the parameters are: $a=2.00$, $b=0.01$, $c=0.75$, $x_0 = -0.75$,
$x_{abs} = -6.0$, $v_{start}=8.62 \times 10^{-5}$.} \label{pf
return}
\end{figure}
In Fig.~\ref{pf volatility} we show the PF of the volatility for our
model, and we can see a log-normal behavior as that observed
approximately in real market data. The agreement is very good for
values of the volatility greater than $v \sim 0.03$ and
discrepancies are in the range of very low volatility values.
Finally in Fig.~\ref{return correlation}a we show the correlation
function of the returns for NLH model and for the real data. The
agreement of the two correlation functions is very good for all the
time. As we can see, the autocorrelations of the asset returns are
insignificant, except for very small time scale for which
microstructure effects come into play. This is in agreement with one
of the stylized empirical facts emerging from the statistical
analysis of price variations in various types of financial markets
[Cont, 2001]. The good agreement between theoretical and
experimental behaviour is confirmed by the correlation function of
the logarithmic absolute returns, which decays slowly to zero (see
Fig.~\ref{return correlation}b).
Our last investigation concerns the PF of the escape time of the
returns, which is the main focus of our paper.
\begin{figure}[htbp]
\vspace{1mm}
\centering{\resizebox{7.5cm}{!}{\includegraphics{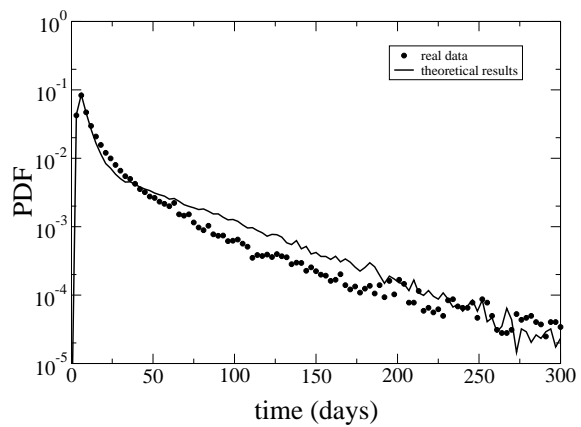}}}
\vskip-0.1cm \caption{Probability function of the escape time of the
returns from simulation (solid line), and from real data (black
circle). The values of the thresholds are: $\Delta x_i = -0.1
\thinspace \sigma_{\Delta x}$, $\Delta x_f = -1.5 \thinspace
\sigma_{\Delta x}$.  The values of the other parameters are the same
as in Fig.~\ref{pf return}} \label{PF}
\end{figure}
By using our model (Eqs.~(7)-(9)), we calculate the probability
function for the escape time of the returns. We define two
thresholds, $\Delta x_i$ and $\Delta x_f$, which represent the start
point and the end point respectively for calculating MET. To fix the
values of the two thresholds we consider the standard deviation (SD)
$\sigma_{\Delta x}$ of the return series over a long time period
corresponding to that of the real data and we set $\Delta x_i = -0.1
\thinspace \sigma_{\Delta x}$, $\Delta x_f = -1.5 \thinspace
\sigma_{\Delta x}$. The initial position is $x_0 = -0.75$ and the
absorbing barrier is at $x_{abs} = -6.0$. We use a trial and error
procedure to select the values of the parameters $a$, $b$, and $c$
for which we obtain the best fitting between all the statistical
features considered, theoretical (NLH model) and empirical (real
data). As real data we use the daily closure prices for $1071$
stocks traded at the NYSE and continuously present in the $12-$year
period $1987-1998$ (3030 trading days). In Fig.~\ref{PF} we report
the results for the PF of the escape times obtained both from real
and theoretical data: we note a good qualitative agreement between
the two PFs. Moreover we check the agreement between the two data
sets by performing both $\chi^2$ and Kolmogorov-Smirnov (K-S)
goodness-of-fit tests. The results are:
$\chi^2\thinspace=\thinspace0.01620$,
$\tilde{\chi}^2\thinspace=\thinspace0.00017$ ($\tilde{\chi}^2$
indicates the reduced $\chi^2$) and $D=0.14$, $P=0.261$, where $D$
and $P$ are respectively the maximum difference between the
cumulative distributions and the corresponding probability for the
K-S test. The results obtained from both tests indicate that the two
distributions of Fig.~\ref{PF} are not significantly different. Of
course, a better quantitative fitting procedure could be done by
considering also the potential parameters. This detailed analysis
will be done in a forthcoming paper.

\section{Conclusions}

We studied the statistical properties of the escape times in
different models for stock market evolution. We compared the PFs of
the escape times of the returns obtained by the basic geometric
Brownian motion model and by two commonly used SV models (GARCH and
Heston models) with the PF of real market data. Our results indeed
show that to fit well the escape time distribution obtained from
market data, it is necessary to take into account the stochasticity
of the volatility. In the nonlinear Heston model, recently
introduced by the authors, after reviewing the role of the CIR
parameters on the dynamics of the model, we analyze in detail the
role of the initial conditions on the escape time from a metastable
state. We found that the NES effect, which could be considered as a
measure of the stabilizing effect of the noise on the marked
dynamics, is more pronounced for unstable initial positions near the
maximum of the potential. For initial positions inside the potential
well we recover an interesting nonmonotonic behavior with a minimum
and a maximum for the MET as a function of the parameter $b$. This
behaviour is a typical signature of the NES effect in the divergent
dynamical regime [Fiasconaro \emph{et al.}, 2005]. To check the
reliability of our NLH model we compare the return correlation
function, the PFs of the returns, the volatility and the escape
times with the corresponding ones obtained from real market data. We
find good agreement for some of these characteristics.

\section*{Acknowledgments}
This work was supported by MIUR and CNISM-INFM.

\end{document}